\documentclass[prl,aps,superscriptaddress,twocolumn]{revtex4}
\usepackage{graphicx}
\usepackage{dcolumn}
\usepackage{color}
\usepackage{amssymb,amsfonts,amsmath}

\usepackage{subfigure}
\begin{document}
% Include your paper's title here
\title{Universal statistics of epithelial tissue topology} 
\author{Abdulaziz~ Abdullah}
\affiliation{Department of Mathematical Sciences, University of Liverpool, Liverpool, UK}
\author{Demetris~ Avraam}
\affiliation{Department of Mathematical Sciences, University of Liverpool, Liverpool, UK}
\author{Oleksandr Chepizhko}
\affiliation{Institut f\"ur Theoretische Physik, Leopold-Franzens-Universit\"at Innsbruck, Technikerstraße 21a, A-6020 Innsbruck, Austria}
\author{Thomas Vaccari}
\affiliation{Department of Biosciences, University of Milan, via Celoria 26, 20133 Milano, Italy}
\author{Stefano Zapperi}
\affiliation{Center for Complexity and Biosystems,Department of Physics, 
University of Milan, Via Celoria 16, 20133 Milano, Italy}
\affiliation{CNR - Consiglio Nazionale delle Ricerche, Istituto di Chimica della Materia Condensata e di Tecnologie per l'Energia, 
Via R. Cozzi 53, 20125 Milano, Italy}
\affiliation{ISI Foundation, Via Chisola 5, Torino, Italy}
\affiliation{Department of Applied Physics, Aalto University, P.O. Box 11100, FIN-00076, Aalto, Finland}
\author{Caterina A. M. La Porta}
\affiliation{Center for Complexity and Biosystems, Department of Environmental Science and Policy, University of Milan, via Celoria 26, 20133 Milano, Italy}
\author{Bakhtier~ Vasiev}
\email{B.Vasiev@liverpool.ac.uk}
\affiliation{Department of Mathematical Sciences, University of Liverpool, Liverpool, UK}

% Place the author information here.  Please hand-code the contact
% information and notecalls; do *not* use \footnote commands.  Let the
% author contact information appear immediately below the author names
% as shown.  We would also prefer that you don't change the type-size
% settings shown here.

% Include the date command, but leave its argument blank.

\begin{abstract}
Cells forming various epithelial tissues have a strikingly universal distribution for the number of their edges. It is generally assumed that this topological feature is predefined by the statistics of individual cell divisions in growing tissue but existing theoretical models are unable to predict the observed distribution. Here we show experimentally, as well as in simulations, that the probability of cellular division increases exponentially with the number of edges of the dividing cell and show analytically that this is responsible for the observed shape of cell-edge distribution.
\end{abstract}
\maketitle

\begin{figure}[htbp]
	\centering
		\includegraphics[width=\columnwidth]{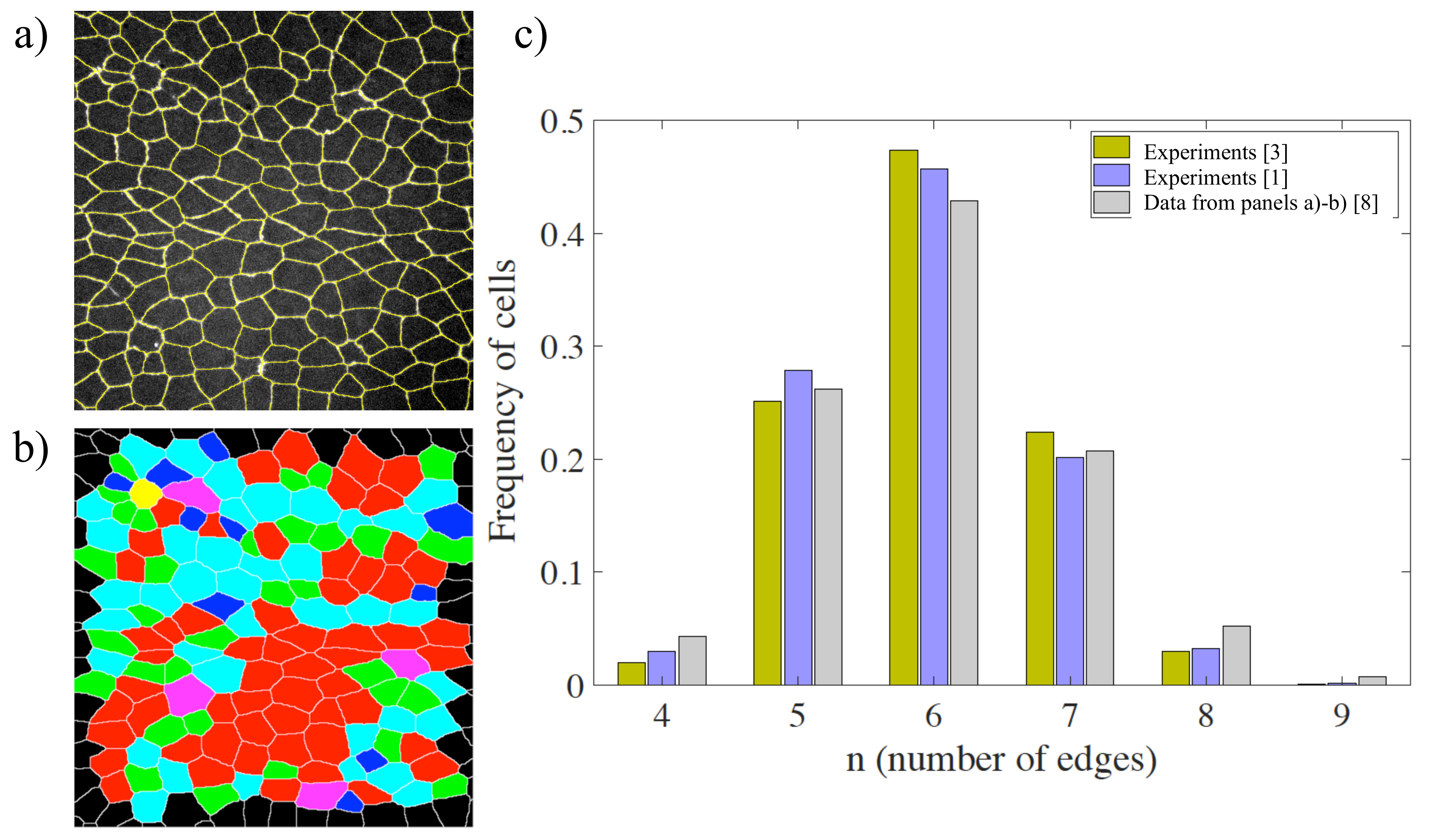}
	\caption{a) A segmented image of the tissue formed by a {\it drosophila} pupa \cite{Besson2015}. b) The
	same image as in a) is colored according to the number of edges in each cell (blue:4, green:5, red:6, cyan:7, yellow:8). c) The corresponding histogram averaged over three different experiments and different time steps. The histograms are compared with earlier results reported in the literature \cite{Gibson2006a,Lewis1926}}
	\label{fig:1}
\end{figure}

Epithelial tissues are commonly represented by unicellular layers and have quite distinctive topological features \cite{Gibson2006a, Lewis1928}. Cells, as seen from the tissue surface, represented by  polygons (Figure \ref{fig:1}a) whose number of edges (or the number of neighboring cells) usually varies between $4$ and $9$
(Figure \ref{fig:1}b). Triangular cells or cells with ten or more edges are rarely met. Histograms displaying the fraction of cells with a given number of edges, cell-edges distribution histograms (CEDH), are commonly used to describe the topology of epithelial tissue \cite{Lewis1926, Sahlin2010} and display a remarkable degree of universality across experiments and species \cite{Sandersius2011a}. CEDH indicates that the majority of the cells in an epithelial tissue have hexagonal topology ($\sim45\%$), pentagonal and heptagonal cells are observed less frequently ($\sim25\%$ and $\sim20\%$ respectively) while $4$-, $8$- and $9$-sided cells are rarely observed with a frequency of less than $10\%$ in total (Figure \ref{fig:1}c). 

A few mathematical models have been developed to explain the universal shape of the observed CEDHs. The model introduced by Gibson et al. \cite{Gibson2006a} (thereafter referred to as the GPNP model), considers cellular proliferation as the sole process responsible for tissue topology. Hence, the number of cell edges is defined when the cell is born, does not change during its growth, but can be affected by the division of neighboring cells. According to the GPNP model, cells are polygons with four or more edges and are divided synchronously in discrete generations, ignoring any spatial correlation between the number of edges of the neighboring cells. The GPNP model reproduces fairly well the observed CEDHs with the only exception that $4$-sided cells do completely disappear. 
Sandersius et al. \cite{Sandersius2011a} attempted to revisit the GPNP model to overcome the shortcoming with the $4$-sided cells. 
They have developed four modifications of the GPNP model but none of them was successful: it turned out that all considered modifications of the GPNP destroy the shape of CEDH and it becomes not comparable with the CEDHs from experiments. The authors concluded that their models fail because they do not take into account spatial correlations between sidedness of neighbouring cells (so called Aboav's law \cite{Chiu1995}) which should play a significant role on the development of a proliferating tissue. They confirmed this conclusion by showing a good agreement between experimental observations and simulations of a computational subcellular element model \cite{Sandersius2011b} although the analytical confirmation was omitted.

\begin{figure}[htbp]
\centering
	\includegraphics[width=\columnwidth]{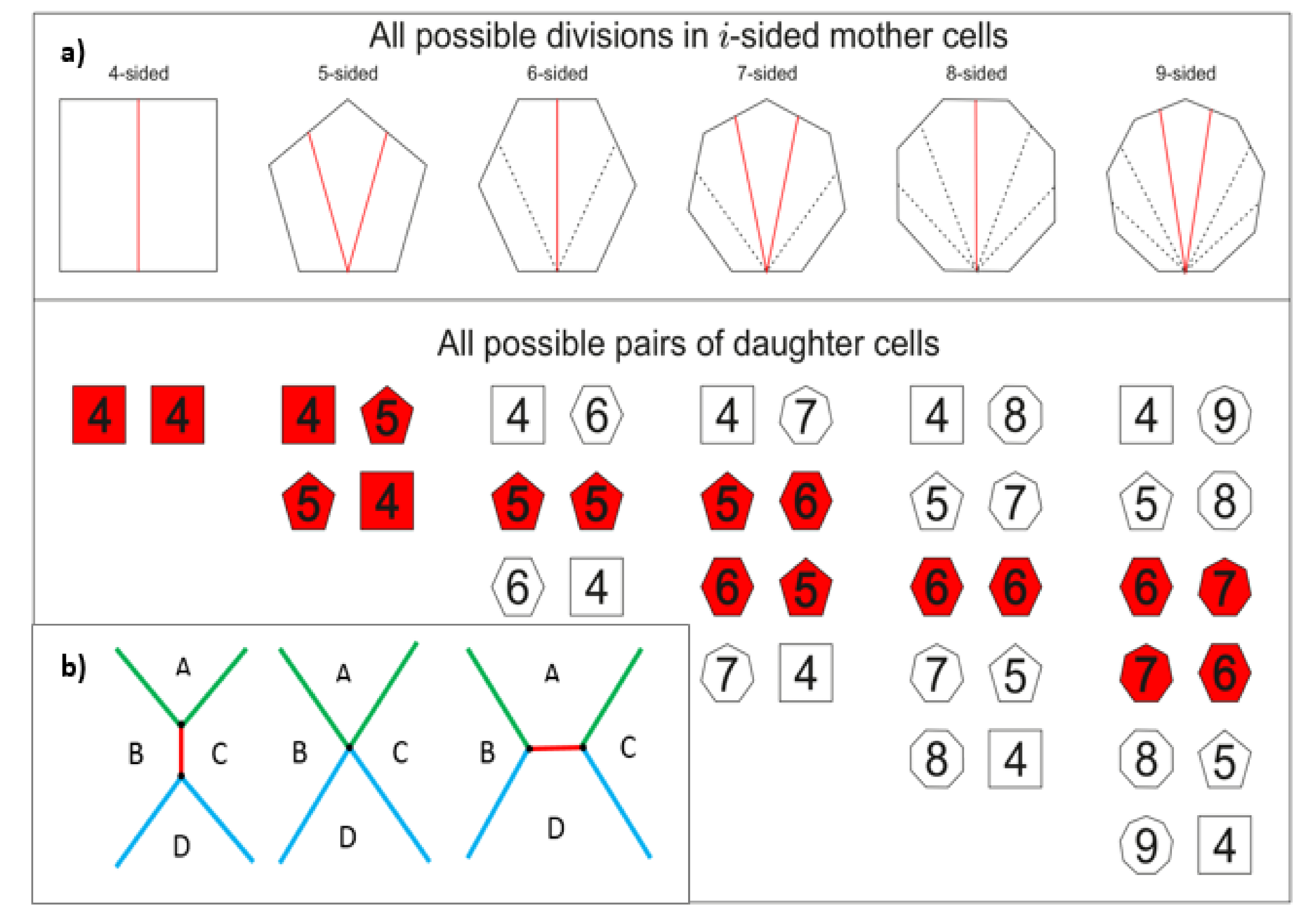}
\caption{a) All possible divisions in $i$-sided mother cells and the corresponding combinations of daughter cells. The red lines in the upper panel and the red colored pairs of daughter cells at the lower panel denote the division patterns and the possible pairs of daughter cells under the `equal split division' scenario \cite{Patel2009}. 
b) Three consecutive images illustrating a T1-transition in dynamic tissue: the edge AD disappears, while the new BC edge forms.}
			\label{fig:2}
		\end{figure}

In this letter, we propose mathematical models based on master equations that describe the evolution of the CEDH due to cellular proliferation and changes in cell shapes. We use these models to identify key features of tissue dynamics responsible for the experimentally observed CEDH.   We show that in order to be successful, a model of growing tissue should account for the fact that cells with more edges divide more frequently. This fact is fully justified: cells maturate before dividing and maturated cells have usually more edges. Using experiments on {\it drosophila} and numerical simulations on a dynamic vertex model, we find that the frequency of cellular division increases exponentially with the number of cell edges. Furthermore, we model the effect of cellular shape changes (due to T1-transitions as illustrated in Figure \ref{fig:2}b) and show that this also allows to reproduce the universal cell-edge distribution. 

For experimental comparison, we consider time lapse images obtained during the development of {\it drosophila} pupae (see Ref. \cite{Besson2015} for experimental details). The determination of the number of edges is done by a custom Matlab code. Initial image segmentation is performed by a watershed algorithm with some preliminary adjustments of the image brightness. Then image segmentation is improved manually (see Figure \ref{fig:1}a). We identify the region corresponding to each cell and count its neighbors (see Figure \ref{fig:1}a). Cells touching the boundary of the image are excluded from the analysis. In this way, we are able to obtain the CEDH (see Figure \ref{fig:1}c) which is in agreement with earlier observations \cite{Gibson2006a,Sandersius2011a}.

In our first model we consider a growing tissue where the CEDH is only affected by cellular proliferation while growing cells do not change their shapes
as in Ref. \cite{Gibson2006a}. To write the master equation, we denote by $N(t)$ the total number of cells at time t; $N_i(t)$- the number of $i$-sided cells, and $p_i(t)=N_i(t)/N(t)$- the fraction of $i$-sided cells in the population. We also assume that the number of edges cells can have varies from $4$ to $9$. The rate of change in the fraction of $i$-sided cells is given by
$$
\frac{d}{dt}\left(\frac{N_{i}}{N}\right)=\frac{dN_{i}N-N_{i}dN}{N^2dt}=\frac{dN}{Ndt}\left(\frac{dN_{i}}{dN}-\frac{N_{i}}{N}\right),
$$
\noindent which can be written as 
\begin{equation}{\label{basic form}}
   \dot{p_{i}}=\alpha\left(M_{i}+K_{i}-p_{i}\right),
   \end{equation}
\noindent where $\alpha=\frac{dN}{Ndt}$ is the cells proliferation rate and the expression in brackets defines the probability for $i$-sided cell to appear/disappear in a single proliferation event. $\frac{dN_{i}}{dN}=M_{i}+K_{i}$ is split into two terms, where $K_{i}$ determines the changes due to removal of $i$-sided mother cells and addition of $i$-sided daughter cells while $K_{i}$ accounts for the changes in the number of edges of neighboring cells after each division. The term $K_{i}$ is easy to estimate assuming that the fraction of $i$-sided cells in the neighborhood of any dividing cell is equal to their total fraction (i.e. there is no spatial correlation). Then:  
   \begin{equation}{\label{TermK}}
			K_{i}=-2p_{i}+2p_{i-1},
	\end{equation}

\noindent where the first term defines the decrease in the fraction of $i$-sided cells if either or both affected neighbours were $i$-sided before the division and become $i+1$-sided after the division, and the second term counts the cells that were $i-1$-sided before the division and become $i$-sided. If we allow for the number of edges to vary from four to nine, then Eq. (\ref{TermK}) holds for $5<i<8$, while for the boundary cases we have $K_{4}=-2p_{4}$ and $K_{9}=2p_{8}$.

The term $M_{i}$ in Eq. (\ref{TermK}) describes changes associated with the replacement of a mother cell by two daughter cells during cell division. In the following, we only consider division events (as listed in \cite{Patel2009}) where the mother cell is split into two approximately equal daughter cells. Hence, we assume that a mother cell with an even number of edges divides into two daughter cells with equal number of edges, while a mother cell with an odd number of edges splits into cells that differ by one in their number of edges. Allowed divisions are illustrated in red in the upper panel of Figure \ref{fig:2}a and possible pairs of daughter cells - in red in the lower panel of Figure \ref{fig:2}a. It is easy to see that $i$-sided daughter cells can only appear after the division of ($2i-3$), ($2i-4$) or ($2i-5$)-sided cells. Then, the term $M_{i}$ can be represented as:
		\begin{equation}{\label{EQSPLIT}}
	M_{i}= 
	\begin{cases}
		2p_{2i-4}^\ast+p_{2i-3}^\ast -p_{i}^\ast							&\quad\text{if}~ i=4\\
		p_{2i-5}^\ast+2p_{2i-4}^\ast+p_{2i-3}^\ast-p_{i}^\ast &\quad\text{if}~4<i<7\\
		p_{2i-5}^\ast-p_{i}^\ast		        									&\quad\text{if}~ i=7\\
		-p_{i}^\ast 										  										&\quad\text{if}~ i>7\\
		\end{cases}
		\end{equation}
				
\begin{figure*}[htbp]
	\centering
	\includegraphics[width=18cm]{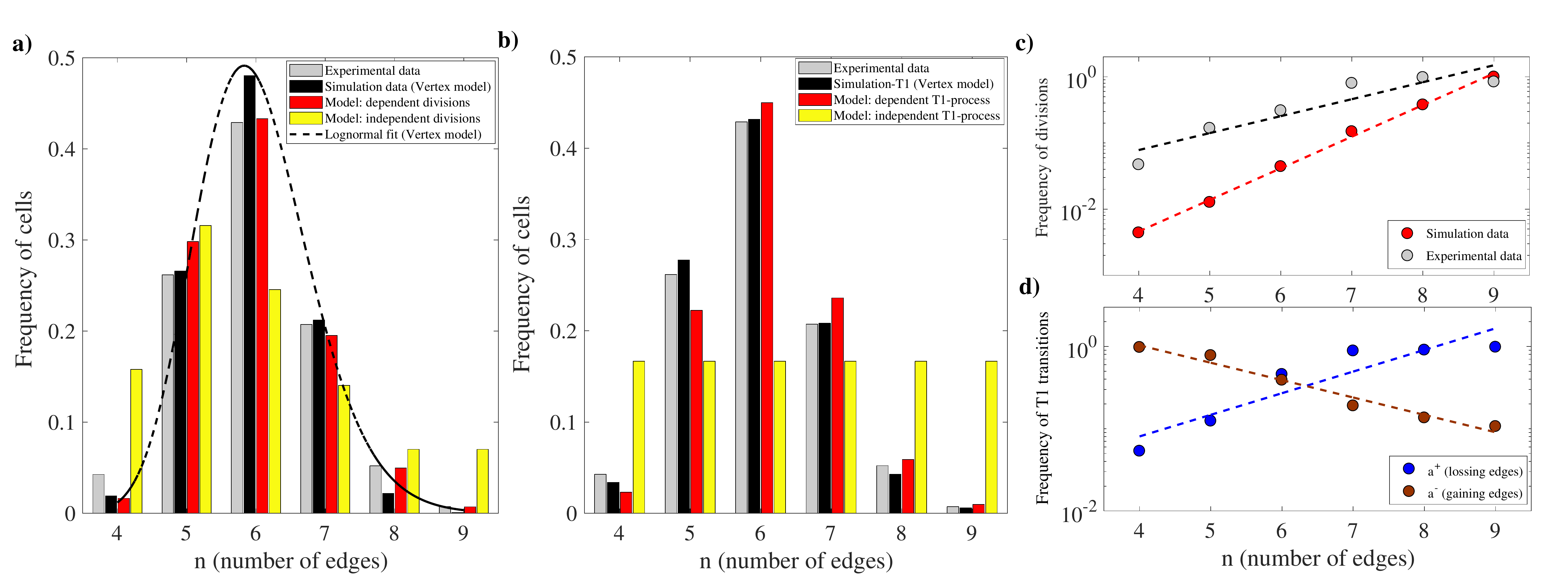}
	\caption{a) CEDHs from simulations and analytical models (neglecting/considering dependance of the probability of cellular division on the number of edges) of growing tissue (compared with experimental CEDH).  
	b) CEDHs from simulations and analytical models (neglecting/considering dependance of the probability of T1-transition on the number of edges) of dynamic tissue (compared with experimental CEDH).
	c) Fractions, $\sigma_n$, of dividing $n$-sided cells as estimated from the experimental images (data from \cite{Besson2015} analyzed here) and from numerical simulations using the vertex model. In both cases, the probability of division grows exponentially with the number of cell edges. 
	d) Fractions of $n$-sided cells gaining ($a_n^+$) and loosing ($a_n^-$) an edge in course of T1-transitions as obtained from numerical simulations using the vertex model. }
	\label{fig:3}
\end{figure*}

\noindent where $p_i^\ast$ is the probability that a mother cell, in a randomly chosen division, is $i$-sided. If, following \cite{Gibson2006a}, we assume that all cells divide with equal probability, i.e. $p_i^\ast=p_i$, then Eqs. (\ref{basic form}-\ref{EQSPLIT}), form a system of linear equations satisfying the condition $\sum_{i=4}^{9}\dot{p_i}=0$ and describing the evolution of CEDH. One can show that this system has one zero eigenvalue while all its other eigenvalues are negative so that its solution converges to the eigenvector corresponding to the zero-eigenvalue. This solution does not, however, reproduce the topology of growing tissue obtained experimentally (see Figure \ref{fig:3}a). 

The assumption of a cell division probability that is independent on the number of edges is not supported by experiments \cite{Lewis1928}. We 
show in Figure \ref{fig:3}c that the division probability is increasing exponentially with the number of edges of the dividing cell. This
result is expected considering that  cells have to mature before they divide and 
the number of edges of long-living cell tends to increase 
due to the division of neighboring cells. We thus modify our model assuming that in Eq. (\ref{EQSPLIT}) the probability that a dividing cell is $i$-sided is proportional to 
$\sigma_{i}p_i$ where $\sigma_{i}$ is estimated from experiments  (Figure \ref{fig:3}c),
yielding 
	\begin{equation}{\label{MEQSPLIT}}
	p_i^\ast=\frac{\sigma_{i}p_i}{\sum_i \sigma_{i}p_i},
		\end{equation}
\noindent which makes our model nonlinear. Solutions of this model yield steady-state results for the distribution of cell edges that are now in excellent agreement with experimental observations as shown in Figure \ref{fig:3}a.

As a further confirmation of the theoretical results, we perform numerical simulations using a dynamical vertex model \cite{Honda2015} which is commonly used in modelling epithelial tissues \cite{Farhadifar2007, Fletcher2014}. In this model, each cell is represented by a polygon whose shape can change due to the forces acting on its vertices. Simulations of the vertex model are implemented through the open software Chaste \cite{Mirams2013}, considering the formation of a tissue from a single cell in the course of successive divisions (see Supplementary Movie 1).  The number of edges of each single cell (and the number of edges of its neighbors) can be explicitly specified and the process of division can be fully determined. We find that the CEDHs for simulated growing tissues converge and (when the  number of cells exceeds 1000) acquire the shape shown by the black histogram in Figure \ref{fig:3}a. Simulations also show that the probability of cellular division is exponentially increasing with the number of its edges (see Figure \ref{fig:3}c). Both results are in excellent agreement with the experimental data. 
 
%%%%%%%%%%%%%%%%%%%%%%%%%%%%%%%%%%%%%%%%%%%%%%%%%%%%%%%%%%%%%%%%%%%%%%%%%%%%%%%%%%%%%%%%%%%%%%%%%%%%%%%%%%%%%%%%%%%%%%%%%%%%%%%%%%%%%
In the model we have considered so far, we have assumed that the CEDH is affected only by cellular proliferation while the number of edges in growing daughter cells does not change, unless affected by dividing neighboring cells. Epithelial cells can, however, show dynamical changes in their shapes, manifested by the so-called T1-transitions \cite{Nagai2001} when one edge disappears (bringing together two distant cells) while another appears (separating two neighbouring cells) (see Figure \ref{fig:2}b). We study the effect of 
T1-transitions on CEDH in simulations using a modified version of the dynamic vertex model. Cells in the simulation are not allowed to proliferate but forced to dynamically change and undergo T1-transitions (see Supplementary Movie 2). We note that T1-transitions do not change neither the number of cells nor the total number of cell edges, so that the outcome of these simulations strongly depends on the initial state of modeled tissue. Here we consider a tissue composed of a considerable amount of cells with six edges per cell on average, corresponding
to the experimental case  \cite{Lewis1928}. These simulations show that if we start with a tissue containing more than 100 hexagonal cells the CEDH evolves towards a stationary shape matching the experimental results (see Figure \ref{fig:3}b). 

We can explain this result theoretically by considering the master equation describing the dynamics of CEDH in a tissue without cell division: 
\begin{equation}{\label{MasterEquationForT1process}}
	\dot{p_{i}}=\alpha F_i,
\end{equation}
\noindent where (similarly to Eq.(\ref{basic form})) $\alpha$ defines the rate at which T1-transitions take place and $F_i$ defines the probability of appearance/disappearance of a $i$-sided cell in a single T1-transition event. To find $F_i$, we note that in each T1-transition two cells loose an edge and two other cells gain an extra edge so that  
		%\begin{widetext}
		\begin{equation}{\label{T1processEquation}}
	F_i= 
	\begin{cases}
		-2p_i^++2p_{i+1}^-					&\quad\text{if}~ i=4\\
		2p_{i-1}^+-2p_{i}^--2p_{i}^++2p_{i+1}^- &\quad\text{if}~i=5,..., 8\\
		2p_{i-1}^+-2p_{i}^-		        				&\quad\text{if}~ i=9
		\end{cases}
		\end{equation}
		%\end{widetext}
\noindent where $p_{i}^-$ and $p_{i}^+$ are probabilities for $i$-sided cell to loose or gain an edge in course of a random T1-transition event. If we assume that all cells have the same probability to undergo a T1-transition then $p_{i}^-=p_{i}^+=p_{i}$. In this case,  Eqs (\ref{MasterEquationForT1process}-\ref{T1processEquation}) define a linear system whose solution converges to the eigenvector corresponding to its zero-eigenvalue. It can be shown analytically as well as numerically (by solving the system (\ref{MasterEquationForT1process}-\ref{T1processEquation})) that all components of this eigenvector are equal yielding a uniform CEDH (see Figure \ref{fig:3}b). which obviously doesn't match experimental data. Another reasonable assumption is that any existing edge can undergo $T1$-transitions with the same probability, leading  to the conclusion that cells with more edges should loose edges more frequently. We have found frequencies at which cells with different sidedness gain or loose edges in simulations using the vertex model (see Figure \ref{fig:3}d). The simulations have shown that indeed the probability that the cell gains/looses an edge decreases/increases with the number of its edges. We take this data into account by reconstructing Eq. (\ref{T1processEquation}) with adjusted probabilities for $i$-sided cells to gain/loose an edge:  
	\begin{equation}{\label{MEQSPLIT1}}
	p_i^+= \frac{a_i^+p_i}{\Sigma a_i^+p_i} \hspace{1 cm}
	p_i^-= \frac{a_i^-p_i}{\Sigma a_i^-p_i},
		\end{equation}
\noindent where the values of $a_i^+$ and $a_i^-$ are taken from the plot in Figure \ref{fig:3}d. The histogram representing the stationary solution of this model is now in a good agreement with experimental data (see Figure \ref{fig:3}b). 
		
In this study, we have addressed the universal topological features of epithelial tissues represented by the shape of the CEDH. This shape (although only comprised by 5 bars) fits log-normal rather than normal distribution (see \ref{fig:3}a) and this observation extends its universality to other objects, for example, to the distribution of sizes of crashed stones in iron/gold mines \cite{Kolmogrov1941}. Another important observation is that according to this shape the epithelial cell has in average 6 edges \cite{Lewis1926}. This number also appears in models of growing tissues: each cell division leads to the formation of one extra cell and 6 extra edges, implying that 6 edges per cell should be present in large tissues. This coincidence points to the crucial role of cellular proliferation in CEDH as checked in GPNP model and its modifications \cite{Gibson2006a, Sandersius2011a}. Our results suggest that the reason why these models do not completely succeed in reproducing the observed CEDHs is that they do not take into account correlations between the number of edges of a cell and its division probability. We find instead, experimentally as well as in simulations, that the probability of cell division (i.e. fraction of dividing $i$-sided cells) increases exponentially with the number of cell edges (Figure \ref{fig:3}c). Finally, our numerical as well as analytical studies have shown that the T1-transitions taking place in dynamic tissue also support the formation of universal CEDH.

This work has been supported by the BBSRC grant BB/K002430/1 to BV.

\bibliography{Reference1}

\end{document}